\documentclass[aps, prd, superscriptaddress,nofootinbib]{revtex4}
\usepackage{latexsym}
\usepackage{euscript}
\usepackage{epsfig,amsmath,amssymb}

\def\be{\begin{equation}}
\def\ee{\end{equation}}
\def\bea{\begin{eqnarray}}
\def\eea{\end{eqnarray}}

\long\def\symbolfootnote[#1]#2{\begingroup%
\def\thefootnote{\fnsymbol{footnote}}\footnote[#1]{#2}\endgroup} 

 \large\normalsize
\def\etal{{\it et al.}}

\begin{document}

\title{Gravitating (field theoretical) cosmic (p,q)-superstrings}

\author{Betti Hartmann}
\email{b.hartmann@jacobs-university.de}
\affiliation{ School of Engineering and Science, Jacobs University Bremen,
28759 Bremen, Germany }

\author{Jon Urrestilla}
\email{j.urrestilla@sussex.ac.uk}
\affiliation{Department of Physics and Astronomy, University of Sussex,
  Brighton,
BN1 9QH, United Kingdom }

\date{today}

\begin{abstract}
We study field theoretical models for cosmic (p,q)-superstrings in a curved space-time. We discuss both
string solutions, i.e. solutions with a conical deficit, but also
so-called Melvin solutions, which have a completely different asymptotic behaviour.
We show that globally regular gravitating (p,q)-strings exist only in a finite domain
of the parameter space and study the dependence of the domain of existence
on the parameters in the model. We find that due to the interaction
between strings, the parameter range where string solution exist is wider than
for non-interacting strings. 
\end{abstract}

\maketitle

\section{Introduction}

The recent realization that some extended objects in fundamental string theory \cite{polchinski} can have cosmological size
has triggered a lot of interest both in the fundamental string and the cosmic string community. 
In braneworld scenarios, big quantities of two types of cosmic size ``strings'' (cosmic superstrings) can be produced \cite{majumdar} when bane-antibrane pairs annihilate
each other: fundamental (F) strings and Dirichlet (D) strings, which can be either a one dimensional brane (D1-brane) or a higher dimensional
brane with all but one dimension compactified. These two kinds of strings interact with each other \cite{dvali} forming bound states: $p$ F-strings can interact
with $q$ D-strings to form  (p,q) bound states, resulting in a network of F-, D-  and FD-strings.

Networks of  (p,q) string have been investigated in several papers lately, both analytically and numerically, and both in the realms 
of fundamental strings and  solitonic gauge strings \cite{bulk,saffin,rajantie,salmi,urrestilla}. 
The dynamics of (p,q) network is more complicated than ``usual'' 
(solitonic) cosmic strings: The intercommutation
of solitonic strings is almost always one, whereas in (p,q) strings it is usually less than one, 
and often much less than one.  
Generally, solitonic strings do not 
form bound states, so that 3-way Y junctions do not form, whereas (p,q) strings are known to form stable bound states, and therefore create 
Y junctions. All strings
in a network of ordinary solitonic strings have the same tension, whereas there is a whole range of different tensions in a  (p,q) network.

Properties of ordinary (solitonic) comics strings have been thoroughly studied, and their network properties are better understood than
those of (p,q) strings. They 
are known to be generically formed in supersymmetric hybrid inflation \cite{lyth} and grand unified based inflationary models \cite{jeannerot}. 
There has 
been considerable effort in 
numerically modelling networks to obtain observables; for example, modeling field theoretical cosmic strings to obtain 
CMB power spectra \cite{cmb}.

The gravitational properties of solitonic cosmic strings have also be studied
in detail by minimally coupling the Abelian Higgs model to gravity. Far away from the 
core of the string, the space-time has a deficit angle, i.e. corresponds to
Minkowski space-time minus a wedge \cite{garfinkle}. The deficit angle is in linear
order proportional to the energy per unit length of the string. 
If the vacuum expectation value (vev) of the Higgs field is sufficiently large
(corresponding to very heavy strings that have formed at energies much bigger than the GUT scale), the deficit angle becomes larger than $2\pi$. These solutions
are the so-called ``supermassive strings'' studied in \cite{gl} and possess
a singularity at a maximal value of the radial coordinate, at which the
angular part of the metric vanishes. Interestingly, it was realized
only a few years ago \cite{clv,bl} that both the globally regular string solution
as well as the supermassive solution have ``shadow'' solutions that exist for the same parameter values. The string-like solutions with deficit angle $ < 2\pi$ have a
shadow solution in the form of so-called Melvin solutions, which have a different
asymptotic behaviour of the metric and have higher energies than their string counterparts (and are thus very likely cosmologically not relevant). The supermassive
string solutions on the other hand have shadow solutions of Kasner-type.
Kasner solutions possess also a singularity at some finite radial distance, the difference is that for Kasner solutions, the $tt$-component of the metric vanishes at this maximal radial distance, while the angular part of the metric diverges there.
Since both supermassive as well as Kasner solutions contain space-time singularities
they are surely of limited interest for cosmological applications.

It is partly due to the fact that advanced machinery  exists to study solitonic strings that different authors have tried to capture properties of (p,q) 
string 
networks using field theoretical models. One can construct field theory models that give rise to solitonic strings that share some of the properties
 of cosmic superstrings. For example, 
Abelian Higgs strings can have a reconnexion probability which is less than one when the strings meet at very high velocities \cite{deputter}. 
Also,  type-I Abelian 
Higgs strings
can form bound states (and have strings of different tensions) when they meet at low velocities \cite{bettencourt,salmi}, 
or if a network of (type I) strings with high
winding numbers forms in a thermal phase transition \cite{donaire}. 
In models with SUSY flat directions strings with type I properties can also be formed \cite{pickles}.

A maybe more realistic approximation is given by the models of Saffin \cite{saffin} and Rajantie, Sakellariadou and Stoica \cite{rajantie}, 
where two independent Abelian 
Higgs string models 
are coupled via an interacting potential. The strength with which the strings interact and form bound states can be controlled by the parameters in 
the potential,
giving rise to numerically more feasible and controllable simulations, including network simulations \cite{rajantie,urrestilla}.

In this paper we couple minimally the models of Saffin and Rajantie \etal\ to gravity, and investigate the gravitating properties of this field theoretical approximations of (p,q) cosmic superstrings. In the next section we describe the models and set up the ans\"atze and boundary conditions. In section \ref{NR} we report the numerical solutions found for the models studied, and then conclude.

\section{The model}
The models we are studying are given by the following action:
\begin{equation}
\label{action}
S=\int d^4 x \sqrt{-g} \left( \frac{1}{16\pi G} R + {\cal L}_{m} \right)
\end{equation}
where $R$ is the Ricci scalar and $G$ denotes Newton's constant. The matter Lagrangian
${\cal L}_{m}$ reads:
\begin{equation}
{\cal L}_{m}=D_{\mu} \phi (D^{\mu} \phi)^*-\frac{1}{4} F_{\mu\nu} F^{\mu\nu}
+D_{\mu} \xi (D^{\mu} \xi)^*-\frac{1}{4} H_{\mu\nu} H^{\mu\nu}
-V(\phi,\xi)
\end{equation} 
with the covariant derivatives $D_\mu\phi=\nabla_{\mu}\phi-ie_1 A_{\mu}\phi$,
$D_\mu\xi=\nabla_{\mu}\xi-ie_2 B_{\mu}\xi$
and the
field strength tensors $F_{\mu\nu}=\partial_\mu A_\nu-\partial_\nu A_\mu$, 
$H_{\mu\nu}=\partial_\mu B_\nu-\partial_\nu B_\mu$  of the two U(1) gauge potential $A_{\mu}$, $B_{\mu}$ with coupling constants $e_1$
and $e_2$. The fields 
$\phi$ and $\xi$ are complex scalar fields (Higgs fields).

We will study two different potentials, corresponding to Saffin \cite{saffin} ($V_1$) and Rajantie \etal\ \cite{rajantie} ($V_2$)

\be
V_1(\phi,\xi)=\frac{\lambda_1}{4}\left(\phi\phi^*-\eta^2_1\right)^2
+\frac{\lambda_2}{4}\left(\xi\xi^*-\eta^2_2\right)^2
-\lambda_3\left(\phi\phi^*-\eta^2_1\right)\left(\xi\xi^*-\eta^2_2\right)
\ee
and
\be
V_2(\phi,\xi)=
\frac{\tilde{\lambda}_1}{4}\left(\phi\phi^*-\eta^2_1\right)^2
+\frac{\tilde{\lambda}_2}{4 \eta_1^2}\phi\phi^*\left(\xi\xi^*-\eta^2_2\right)^2
\ee

In the first model\footnote{This type of potential has also been studied in relation to composite topological defects \cite{ahu,composite}.} ($V_1$) the interaction term makes the strings form a  bound state depending on the value of 
$\Delta=\lambda_1\lambda_2-4\lambda_3^2$ \cite{ahu,saffin}: for $0< \Delta < \frac{1}{2} \sqrt{\lambda_1 \lambda_2}$ bounds states will be formed.  In this 
paper we will investigate the special case where the parameters correspond to the Bogomol'nyi limit (more explicitly, to what 
would be the  Bogomol'nyi limit if $\lambda_3=0$ and $G=0$)
\be
\lambda_1=2e_1^2\qquad\lambda_2=2e_2^2
\ee
and we will eventually rescale parameters such that $\eta_1=\eta_2=e_1=e_2=1$, though we will keep them explicit throughout our equations. It is also worth reminding
that by setting $\lambda_3=0$ we recover two non-interacting Abelian Higgs strings
in curved space-time.

For the second case, the interaction ensures that it is energetically favourable to form bound states for $\tilde{\lambda}_1>0$ and $\tilde{\lambda}_2 > 0$ (otherwise, the
potential will not be bounded from below). This is clearly seen by observing that the potential energy of the $\xi$ field vanishes where $\chi=0$, therefore
favouring that both string cores lie on top of each other.

\subsection{The Ansatz}

In the following we shall analyse  the system of coupled
differential equations associated
with the gravitationally coupled system described above. This system will contain the Euler-Lagrange equations
for the matter fields and the Einstein equations for the metric fields. In order to do that,
let us write down the matter and gravitational fields as shown below.
The most general, cylindrically symmetric line element invariant under boosts
along the $z-$direction is:
\begin{equation}
ds^2=N^2(\rho)dt^2-d\rho^2-L^2(\rho)d\varphi^2-N^2(\rho)dz^2 \ .
\end{equation}
The non-vanishing components of the Ricci tensor $R_{\mu}^{\nu}$ then read \cite{clv}:
\begin{eqnarray}
R_0^0=-\frac{(LNN')'}{N^2 L} \ \ , \ \ R_{\rho}^{\rho} = -\frac{2N''}{N}-\frac{L''}{L} \ \ \ , \ \ R_{\varphi}^{\varphi}= -\frac{(N^2 L')'}{N^2 L} \ \ \ , \ \ \ R_z^z=R_0^0
\end{eqnarray}
where the prime denotes the derivative with respect to $\rho$. 

For the matter and gauge fields, we have \cite{no}:
\begin{equation}
\phi(\rho,\varphi)=\eta_1 h(\rho)e^{i n\varphi} \ ,
\end{equation}
\begin{equation}
\xi(\rho,\varphi)=\eta_1 f(\rho)e^{i m\varphi} \ ,
\end{equation}
\begin{equation}
A_{\mu}dx^{\mu}=\frac {1}{e_1}(n-P(\rho)) d\varphi \ .
\end{equation}
\begin{equation}
B_{\mu}dx^{\mu}=\frac {1}{e_2}(m-R(\rho)) d\varphi \ .
\end{equation}
$n$ and $m$ are integers indexing the vorticity of the two Higgs fields  around the $z-$axis.

\subsection{Equations of motion}
We define the following dimensionless variable and function:

\begin{equation}
x=e_1\eta_1 \rho \ \ \ , \ \ \ L(x)= \eta_1 e_1 L(\rho) \ .
\end{equation}

Then, the total Lagrangian only depends on the following dimensionless coupling constants

\begin{equation}
\gamma=8\pi G\eta_1^2 \ \ , \ g=\frac{e_2}{e_1} \ \ , \ 
q=\frac{\eta_2}{\eta_1} \ \ , \ \ 
\beta_i=\frac{\lambda_i}{e_1^2} \ , \ i=1,2,3  \ \ {\rm or} \ \ 
\tilde{\beta}_i=\frac{\tilde{\lambda}_i}{e_1^2} \ , \ i=1,2 \ .
\end{equation}

Varying the action with respect to the matter fields and metric functions, we
obtain a system of 
six non-linear differential equations. The Euler-Lagrange equations for the matter field functions read:
\begin{equation}
\frac{(N^2Lh')'}{N^2L}=\frac{P^2 h}{L^2}+\frac{1}{2} \frac{\partial
  V_i}{\partial h} \ , \ i=1,2
\end{equation}
\begin{equation}
\frac{(N^2Lf')'}{N^2L}=\frac{R^2f}{L^2}+\frac{1}{2} \frac{\partial
  V_i}{\partial f} \ , \ i=1,2
\label{eqf}
\end{equation}
\begin{equation}
\frac{L}{N^2}\left(\frac{N^2P'}{L}\right)'=2 h^2 P \ ,
\end{equation}
\begin{equation}
\frac{L}{N^2}\left(\frac{N^2R'}{L}\right)'=2 g^2 f^2 R \ ,
\end{equation}
where the prime now and in the following denotes the derivative with respect to $x$.

We use the Einstein equations in the following form:
\begin{equation}
R_{\mu\nu}=-\gamma \left(T_{\mu\nu}-\frac{1}{2}g_{\mu\nu} T\right) \ \ , \ \ \mu,\nu=t,x,\varphi,z
\end{equation}
where $T$ is the trace of the energy momentum tensor $T=T^{\lambda}_{\lambda}$ and the
non-vanishing components of the energy-momentum tensor are (we use the notation
of \cite{clv}) with $i=1,2$:
\begin{eqnarray}
(T_0^0)_i &=& \epsilon_s + \epsilon_v + \epsilon_w + u_i  \nonumber \\
(T_x^x)_i &=& -\epsilon_s - \epsilon_v + \epsilon_w + u_i \nonumber \\
(T_{\varphi}^{\varphi})_i &=&\epsilon_s - \epsilon_v - \epsilon_w + u_i  \nonumber \\
(T_z^z)_i &=&  (T_0^0)_i   \nonumber \\
\end{eqnarray}
where
\begin{equation}
\epsilon_s= (h')^2 + (f')^2   \ \ \ , \ \ \ \epsilon_v = \frac{(P')^2}{2 L^2} + \frac{(R')^2}{2 g^2 L^2} \ \ \ , \ \ \ \epsilon_w = \frac{h^2 P^2}{L^2} + \frac{R^2 f^2}{L^2} \end{equation}
and
\begin{eqnarray}
u_1 & = & \frac{\beta_1}{4}\left(h^2-1\right)^2 + \frac{\beta_2}{4} \left(f^2-q^2\right)^2 -
\beta_3\left(h^2-1\right)\left(f^2-q^2\right)  \ \ \ {\rm or} \nonumber \\
u_2  &=&  \frac{\tilde{\beta}_1}{4}\left(h^2-1\right)^2 + \frac{\tilde{\beta}_2}{4} h^2 \left(f^2-q^2\right)^2
\end{eqnarray}
corresponding to the choice of potential $V_1$ and $V_2$, respectively.

We then obtain
\begin{eqnarray}
\label{N1}
\frac{(LNN')'}{N^2 L}&=& \gamma\left[\frac{(P')^2}
{2 L^2}+ \frac{(R')^2}
{2 g^2 L^2}
-u_i\right] \ , \ i=1,2
\end{eqnarray}
and:
\begin{eqnarray}
\label{N2}
\frac{(N^2L')'}{N^2L}&=&-\gamma\left[\frac{2 h^2 P^2}
{L^2}+\frac{2 R^2 f^2}{L^2}+\frac{(P')^2}{2 L^2}+
\frac{(R')^2}{2 g^2 L^2 } + u_i\right]  \ , \ i=1,2
\end{eqnarray}

\subsection{Boundary conditions}
The requirement of regularity at the origin leads to the  following boundary 
conditions:
\begin{equation}
h(0)=0, \ f(0)=0 \ , \ P(0)=n \ , \ R(0)=m
\label{eom1}
\end{equation}
for the matter fields and 
\begin{equation}
\label{zero}
N(0)=1, \ N'(0)=0, \ L(0)=0 \ , \ L'(0)=1 \ .
\end{equation}
for the metric fields. 

In the special case where $n\neq 0$ and $m=0$ 
the boundary conditions (\ref{eom1}) change according to:
\begin{equation}
h(0)=0, \ f'(0)=0 \ , \ P(0)=n \ , \ R(0)=0 \,.
\end{equation}
while for a $n=0$, $m\neq 0$ string, they read
\begin{equation}
h'(0)=0, \ f(0)=0 \ , \ P(0)=0 \ , \ R(0)=m \ .
\end{equation}

By letting the derivatives of the non-winding scalar field 
be zero at the origin instead of imposing the boundary conditions
for the fields themselves, the non-winding scalar field can take non-zero values at the origin and, as it 
is indeed the case, a ``condensate'' of the non-winding scalar field will form in the core of the winding string.

The finiteness of the energy per unit length requires:
\begin{equation}
h(\infty)=1, \ f(\infty)=q \ , \ P(\infty)=0 \ , \ R(\infty)=0  \ .
\end{equation}

We define as inertial mass per unit length of the solution
\begin{equation}
 E_{in}=\int \sqrt{-g_3} T^0_0 dx d\varphi
\end{equation}
where $g_3$ is the determinant of the $2+1$-dimensional space-time given by $(t,x,\varphi)$.
This then reads:
\begin{equation}
 E_{in}=2\pi\int_{0}^{\infty} NL \left(\epsilon_s + \epsilon_v + \epsilon_w + u_i\right)
\end{equation}

As mentioned earlier, for Saffin's model $V_1$, $\beta_3=0$ corresponds to two non-interacting strings. In the BPS limit 
(the limit we are dealing with) the inertial mass (in units of $2\pi$)
of the $(1,1)$ string is just $E_{in}(\gamma)=2$. This is because the mass is equal
to the sum of the masses of the $(1,0)$-string and of the $(0,1)$-string, which is
$E_{in}^{(1,0)}(\gamma)=E_{in}^{(0,1)}(\gamma)=1$. Correspondingly, the deficit angle
(in units of $2\pi$) is just $\delta/(2\pi)=2\gamma$.

We can then define the binding energy of an $(n,m)$ string as
\begin{equation}
 E^{(n,m)}_{bin}=E_{in}^{(n,m)}-n E_{in}^{(1,0)}-m E_{in}^{(0,1)}
\end{equation}

\section{Numerical results}
\label{NR}
\subsection{Generalities}
The solutions for $\gamma=0$, i.e. in flat space-time have been discussed
in \cite{saffin} and \cite{rajantie}. When $\gamma$ is increased from zero, the (p,q)-strings
get deformed by gravity. If $\gamma$ is not too large, globally regular solutions
exist. These solutions fall into two categories \cite{clv,bl}: the string solutions and the Melvin solutions, where the latter have no flat-space counterpart.
The string solutions are those of astrophysical interest since they describe solutions
with a deficit angle and because the Melvin solutions are heavier than their string counterparts. The metric functions then have the following behaviour at
infinity:
\begin{equation}
 N(x\rightarrow \infty) \rightarrow c_1  \ \ , \ \  L(x\rightarrow \infty) \rightarrow c_2x + c_3 \ \ , \ \ c_2 > 0 \ ,
\end{equation}
where the $c_i$, $i=1,2,3$ are constants that depend on the choice of $\gamma$, $\beta_i$, $i=1,2,3$, $q$ and $g$. The solution has a deficit
angle which is given by 
\begin{equation}
 \delta = 2\pi(1-c_2)
\end{equation}
If the deficit angle $\delta > 2\pi$, $c_2 < 0$ and the solution would have a 
singularity at a finite, parameter-dependent value of $x=x_0$ with $L(x=x_0)=0$, while
$N(x_0)$ stays finite. These solutions are the so-called supermassive string solutions \cite{gl} (or inverted string solutions).

The Melvin solutions exist for the same parameter values as the string solutions, but have a different asymptotic behaviour:
\begin{equation}
 N(x\rightarrow \infty) \rightarrow a_1 x^{2/3}  \ \ , \ \  L(x\rightarrow \infty) \rightarrow a_2 x^{-1/3} \ \ ,
\end{equation}
where again $a_1$ and $a_2$ are parameter dependent constants. This latter solution
is thus not a cosmic string solution in the standard sense since it corresponds
to a solution for which the circumference of circles will decay in the asymptotic
region.

\begin{figure}
\includegraphics[height=6cm,width=7cm]{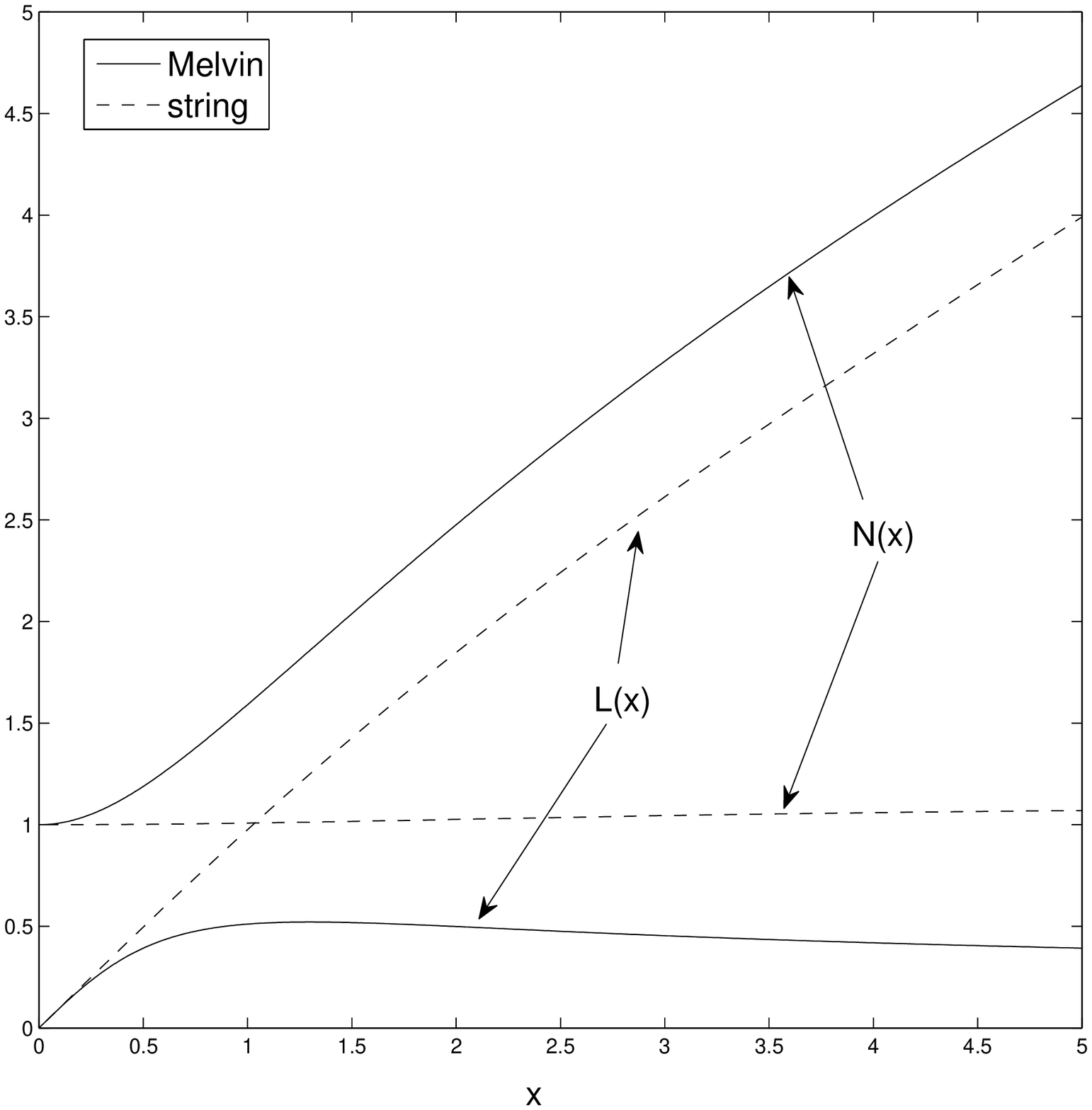} 
\includegraphics[height=6cm,width=7cm]{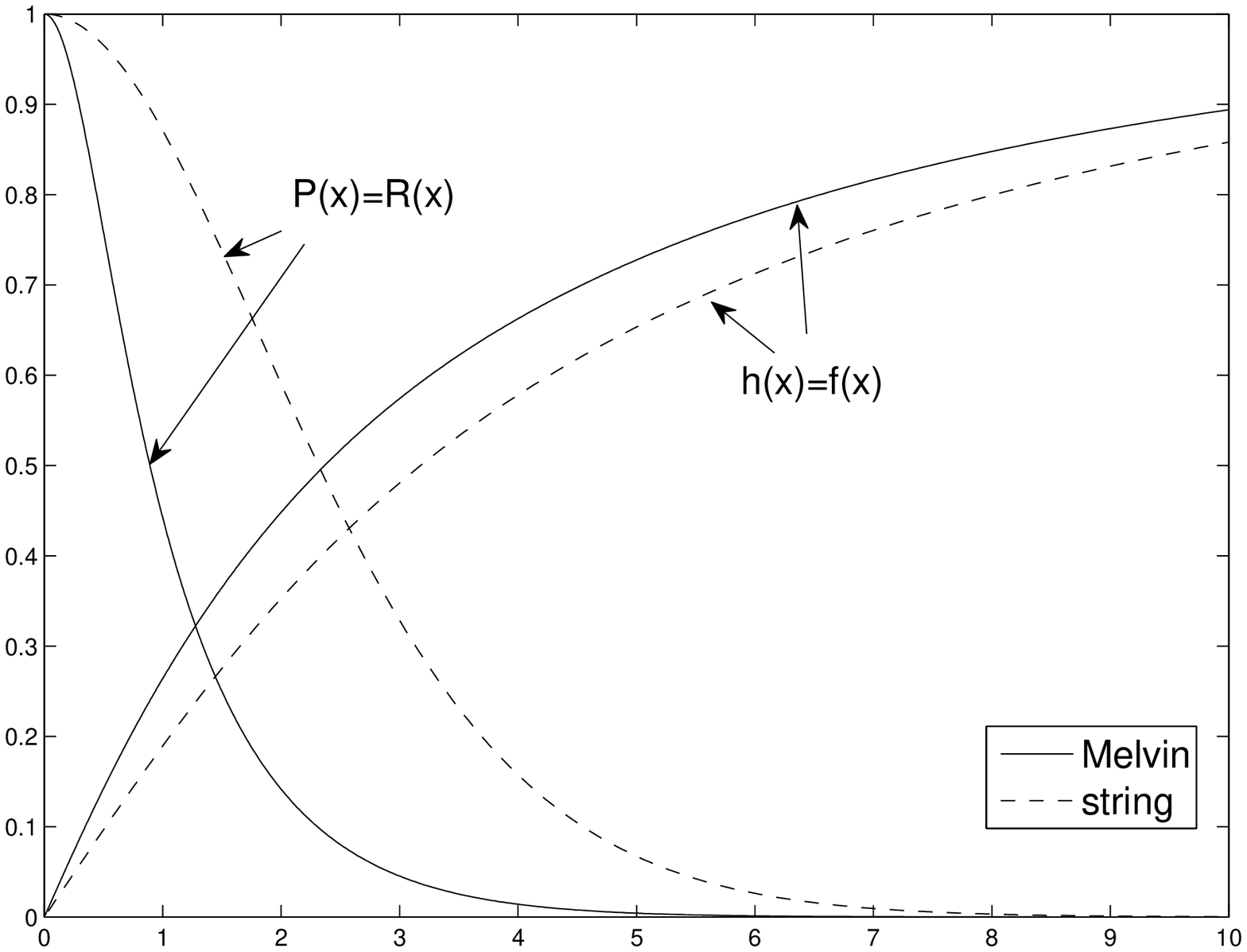}
\caption{\label{fig0} The profiles of the metric functions $N(x)$ and $L(x)$ (left) and
of the matter functions $P(x)=R(x)$, $h(x)=f(x)$ (right) are shown for
a Melvin solution (solid) and a string solution (dashed). Here $\beta_1=\beta_2=2$,
$\beta_3=0.99$, $q=g=1$, $\gamma=0.5$ and $n=m=1$.  }
\end{figure}

The profiles of typical string and Melvin solutions are shown in Fig.~\ref{fig0}.

\subsection{String solutions for the potential $V_1$}
We have solved the above system of equations for $\beta_i=2$, $i=1,2$, $g=1$ ( which for $\gamma=0$ corresponds to the BPS limit) and $q=1$. First, we have fixed $\gamma=0.5$ and investigated the dependence of the deficit angle on the binding parameter $\beta_3$ and on the choice of $n$ and $m$. Note that in order for
$h(x)=1$, $f(x)=1$ to be the global minimum of the theory for our choice
of the remaining parameters given above, we have to require
$0<\beta_3 < 1$. 

\begin{figure}[!htb]
\centering
\includegraphics[width=11cm]{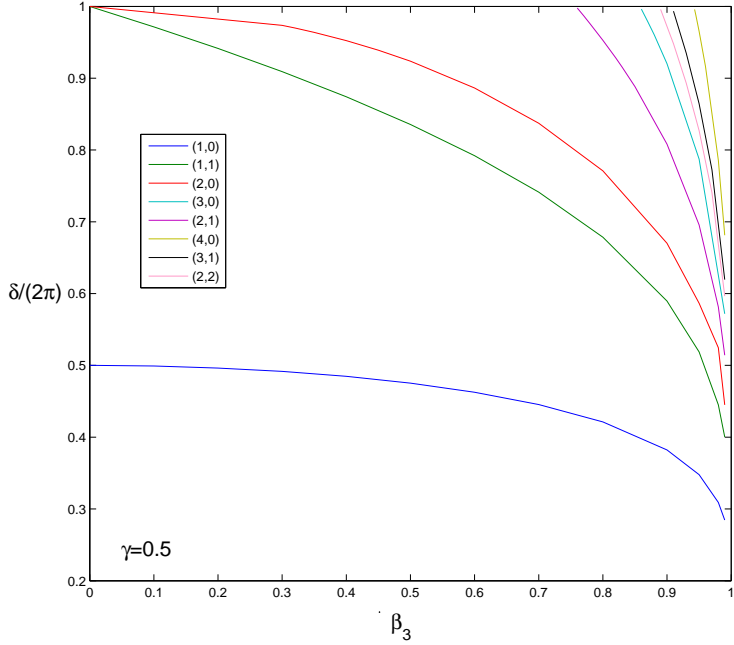} 
\caption{The value of the deficit angle $\delta$ (in units of $2\pi$) is shown as function of $\beta_3$ for $\gamma=0.5$ and different values of $(n,m)$.
 }
\label{fig1}
\end{figure}

Our numerical results are given in Fig.~\ref{fig1}. The deficit angle is decreasing for increasing $\beta_3$ and
tends to zero in the limit $\beta_3\rightarrow 1$. This is expected, since the increase of $\beta_3$ leads
to a stronger binding between the strings, the total energy decreases and thus results
in a smaller deficit angle. Moreover, the deficit angle for a fixed value of $\beta_3$ is smaller for an $(n,n)$ string than for an $(n+1,n-1)$ string, which itself is smaller
than that for an $(n+2,n-2)$ string etc. This is clearly seen for the deficit
angle of the $(2,0)$ string in comparison to that of the $(1,1)$ string and accordingly
for strings with $n+m=3$ and $n+m=4$, respectively.

It is also obvious from Fig.~\ref{fig1} that globally regular string solutions exist only
for a specific parameter range. For $\beta_3=0$, the deficit angle is given by
$\delta/(2\pi)=(n+m)\gamma$. Since globally regular string solutions exist
only for $\delta \leq 2\pi$, we need $(n+m)\gamma \leq 1$ or for $\gamma=0.5$, $n+m \leq 2$. Thus, for our specific choice of $\gamma$ and for $\beta_3=0$ we would not expect regular string solutions to exist for $n+m \geq 3$. This is clearly seen in the figure,
where string solutions with deficit angle less than $2\pi$ exist for all
values of $\beta_3$ only for the $(1,0)$, $(1,1)$ and $(2,0)$ cases. 
However, regular string solutions exist also for $n+m \geq 3$ if $\beta_3$ is large
enough. For $n+m \geq 3$ and a fixed value of $\gamma$, regular string
solutions exist for $\beta_3 > \beta_3^{(cr)}(\gamma,n,m)$. We find e.g.
$\beta_3^{(cr)}(0.5,2,1)\approx 0.76$, $\beta_3^{(cr)}(0.5,3,0)\approx 0.86$,
$\beta_3^{(cr)}(0.5,2,2)\approx 0.89$ and $\beta_3^{(cr)}(0.5,4,0)\approx 0.94$.
The value $\beta_3^{(cr)}$ increases for $n+m$ increasing, since a higher total winding requires a higher binding energy in order to render the string not too heavy. Moreover, for $(n,m)=(n+k,n-k)$ strings, where $0 \leq k \leq n$, the value of $\beta_3^{(cr)}$ increases for increasing $k$.

\begin{figure}[!htbp]
\centering
    \includegraphics[width=11cm]{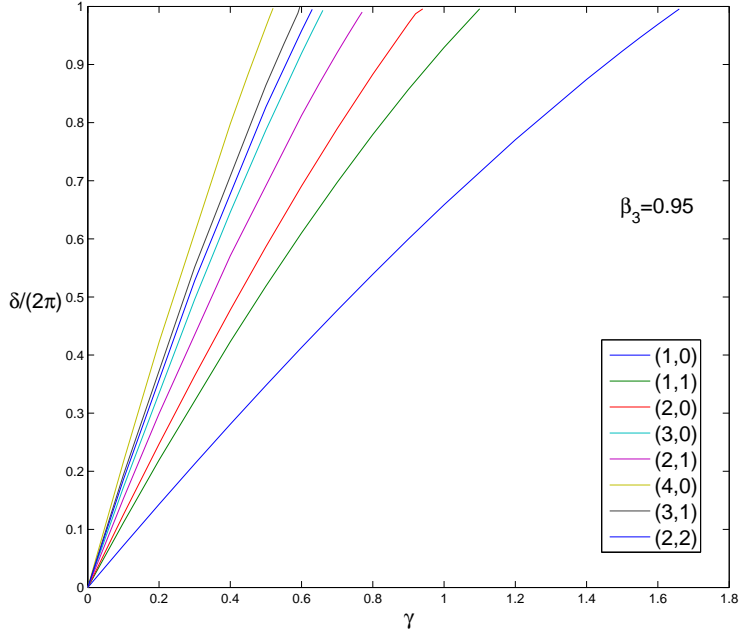} 
\caption{The value of the deficit angle $\delta$ (in units of $2\pi$) is shown as function of $\gamma$ for $\beta_3=0.95$ and different values of $(n,m)$.
 }
\label{fig2}
\end{figure}

We have also studied the dependence of the deficit angle on the gravitational
coupling for a fixed value of $\beta_3=0.95$. The results are given in Fig.~\ref{fig2}.

As expected the deficit angle increases for increasing $\gamma$ and tends to
$2\pi$ for $\gamma\rightarrow \gamma^{(cr)}(\beta_3,n,m)$. Thus, globally regular
strings exist only for $\gamma \leq \gamma^{(cr)}(\beta_3,n,m)$.
This critical value of the gravitational coupling decreases for $n+m$ increasing
and for a fixed value of $n+m$ and $(n,m)=(n+k,n-k)$, $0 \leq k \leq n$, decreases for $k$ increasing.

\begin{figure}[!htbp]
\centering
    \includegraphics[width=11cm]{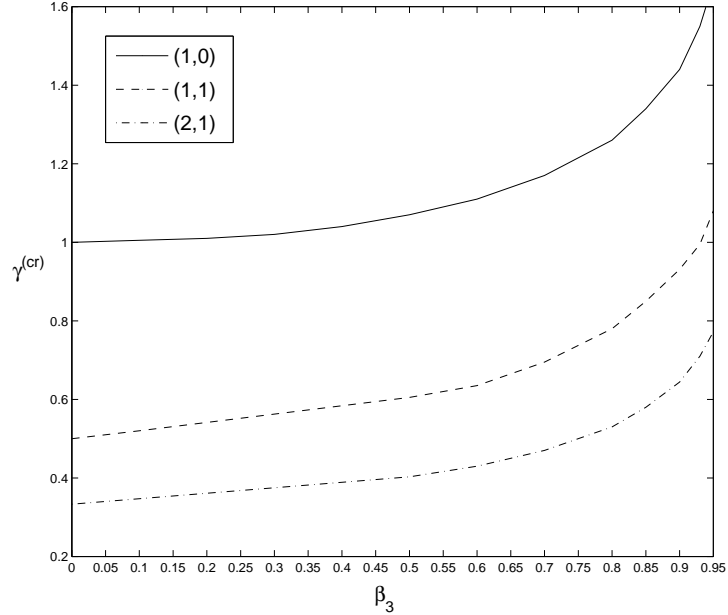} 
\caption{The value of $\gamma^{(cr)}$ is shown as function of $\beta_3$ for
$(1,0)$, $(1,1)$ and $(2,1)$ strings. Note that globally regular
solutions for the specific choices of $(n,m)$ exist only in the parameter domain
below the corresponding curves.}
\label{fig_gamma_cr}
\end{figure}

Clearly, the specific choice of $\beta_3$, $\gamma$ and $(n,m)$ determines
whether globally regular solutions with deficit angle smaller than $2\pi$ exist.
Increasing $\beta_3$ leads to a stronger binding between the strings, i.e.
the energy of the $(n,m)$-string is lowered and hence globally regular
solutions exist for higher values of $\gamma$. This is demonstrated 
in Fig.~\ref{fig_gamma_cr} for $(1,0)$, $(1,1)$ and $(2,1)$ strings. Here, we plot
the value of the gravitational coupling up to which the regular solutions
exist, $\gamma^{(cr)}$, as function of $\beta_3$. Note that regular
solutions exist only in the parameter domain below the respective curves. 
It is clear that
the solutions exist for higher values of $\gamma$ when $\beta_3$ is increased.
This is related to the increase in binding energy, e.g. for the $(2,1)$ solution with
$\gamma=0.5$, we
find that $E_{bin}^{(2,1)}\approx -0.551$ for $\beta_3=0.8$, while
$E_{bin}^{(2,1)}\approx -0.652$ for $\beta_3=0.95$. In addition to the binding
due to the potential, there is an additional binding due to
gravity. The larger $\gamma$, the stronger is the binding: e.g. for $\beta_3=0.9$, we find  $E_{bin}^{(2,1)}\approx -0.621$ for $\gamma=0.5$, while $E_{bin}^{(2,1)}\approx -0.581$ for $\gamma=0$, in agreement with what it has been reported \cite{saffin}.

\subsection{String solutions for the potential $V_2$}

First, we have studied the solutions for $\tilde{\beta}_1=\tilde{\beta}_2=1$, $q=g=1$
and $\gamma=0.5$ to understand what qualitative differences appear in comparison
to the potential $V_1$. While in the case of the $V_1$ potential (and in the BPS limit)
a $(1,0)$-string is exactly equivalent to a $(0,1)$ string, this is different
here. As in the $V_1$ case, when one string is winding and the other is not, the non-winding string 
can form a condensate at the core of the other string. However, in this case, a condensate of the $\xi$ 
field in the core of a $\phi$ string does not lower the energy; whereas a condensate of the $\phi$ inside the $\xi$ core does. More specifically, for $n=1$, $m=0$, the scalar field function $f(x)$ 
is not forced to zero at
the origin and could develop a condensate. However, $f(x)\equiv q$ everywhere
and correspondingly $R(x)=0$ gives the minimum for the potential for the boundary conditions we have.
On the other hand, for $n=0$ and $m=1$, the scalar
field function $h(x)$ is also not forced to zero at the origin, but  does develop a condensate.
At the origin, a value of $h(x)<1$ is the value that minimizes the energy
best. Thus, a $(1,0)$-string is different
from a $(0,1)$ string. This can also be seen when looking at the numerical values of the energies
and deficit angles. For $(1,0)$, we find $E_{in}\approx 0.859$ and $\delta/(2\pi)\approx 0.425$, while for $(0,1)$, we have $E_{in}\approx 0.848$ and $\delta/(2\pi)\approx 0.418$ (see also Fig.~\ref{pot2_critical}).

\begin{figure}[!htbp]
\centering
    \includegraphics[width=11cm]{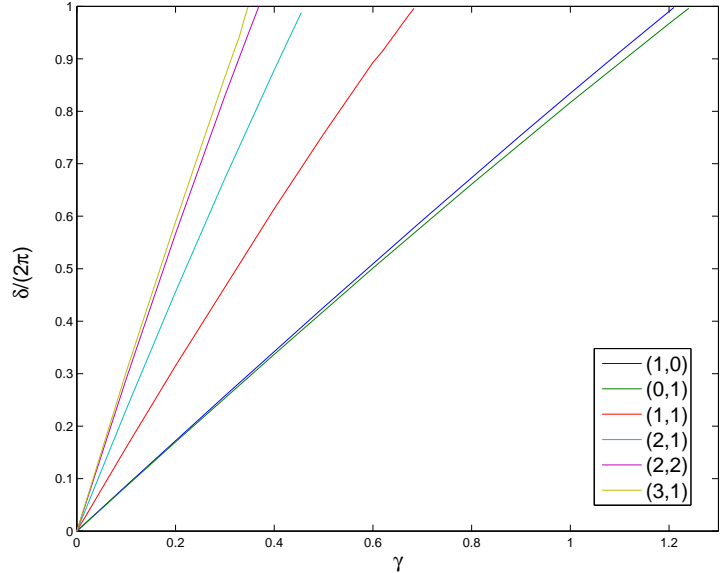} 
\caption{The value of the deficit angle $\delta$ (in units of $2\pi$) is shown as function of $\gamma$ for $\tilde{\beta}_1=\tilde{\beta_2}=1$, $q=g=1$ and different values of $(n,m)$.
 }
\label{pot2}
\end{figure}

In Fig.~\ref{pot2} we show the dependence of the deficit angle on the gravitational coupling $\gamma$. This is qualitatively very similar to the case with the $V_1$ potential. Again, we observe that globally regular string like solutions exist
only up to a maximal value of the gravitational coupling $\gamma=\gamma^{(cr)}(n,m)$. If one would increase
$\gamma$ further, the solutions would have a deficit angle $> 2\pi$ and would thus be
singular. 
 $\gamma^{(cr)}(n,m)$  decreases for $n+m$ increasing
and for a fixed value of $n+m$ and $(n,m)=(n+k,n-k)$, $0 \leq k \leq n$, decreases for $k$ increasing.

As for the $V_1$ case, the specific choice of $\beta_3$, 
$\gamma$ and $(n,m)$ determines, whether globally regular solutions exist.
The fact that the strings interact
and form bound states, lowers their energy. Therefore, for values of $\gamma$ and $(n,m)$ where one would not 
expect regular global solution in the non-interacting case, regular solutions
 exists when strings interact.

\subsection{Melvin solutions}
We have also studied the corresponding Melvin solutions for both potentials. Like their string counterparts, 
they exist only in a limited domain of the parameter plane.

\begin{figure}[!htbp]
\centering
    \includegraphics[width=11cm]{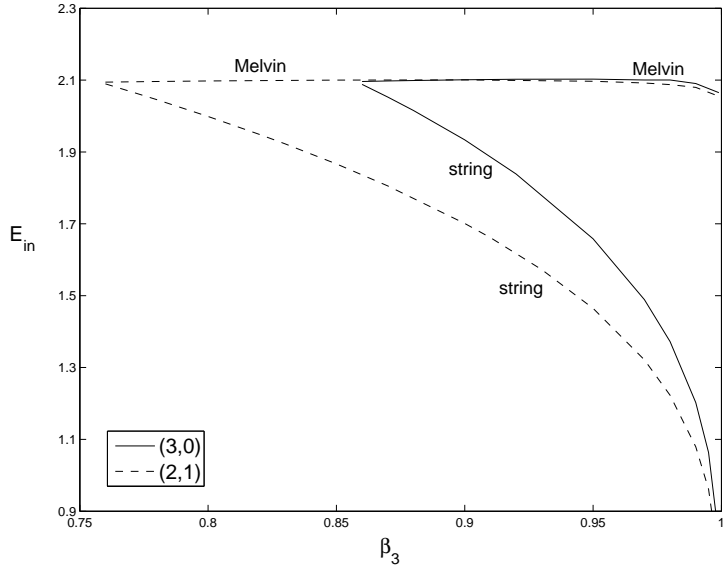} 
\caption{The dependence of the inertial energy $E_{in}$ on the binding
parameter $\beta_3$ is shown for string and Melvin solutions
with $(n,m)=(3,0)$ (solid) and $(n,m)=(2,1)$ (dashed). Here $\beta_1=\beta_2=2$, $q=g=1$
and $\gamma=0.5$. 
 }
\label{fig4}
\end{figure}

In the  $V_1$ case, for a fixed value of $\gamma$, $\beta_3^{(cr)}$ is equal for the string and
Melvin solutions. This has already been observed in the 
case of a single gravitating string \cite{clv,bl}.
At this $\beta_3^{(cr)}$, the inertial energies $E_{in}$ of the two types of solutions
become equal. This is shown in Fig.~\ref{fig4} for $(n,m)=(3,0)$ and
$(n,m)=(2,1)$, respectively. At the critical value of
$\beta_3^{(cr)}$ the curves of the inertial energies of
the two types of solutions merge. Note also that for all cases studied in this
paper, the inertial energy of the Melvin solutions is larger than that of the
corresponding string solutions.

For $\beta_3 < \beta_3^{(cr)}$, the string solutions become
inverted/supermassive string solutions with a zero of the function $L(x)$ at some parameter
dependent value of $x=x_0$, while $N(x=x_0)$ stays finite. The Melvin solutions - on the other hand - 
become so-called Kasner solutions with  $N(x=\tilde{x}_0)=0$ and $L(x\rightarrow\tilde{x}_0)\rightarrow \infty$ at some finite and parameter-dependent
value of $x=\tilde{x}_0$. The inverted string and Kasner solutions
thus exist only on a finite range of the coordinate $x$ and are thus sometimes
also called ``closed'' solutions. 
These solutions have singularities in their metric, and it is out of the scope of this paper to investigate their implications.

\begin{figure}[!htbp]
\centering
    \includegraphics[width=11cm]{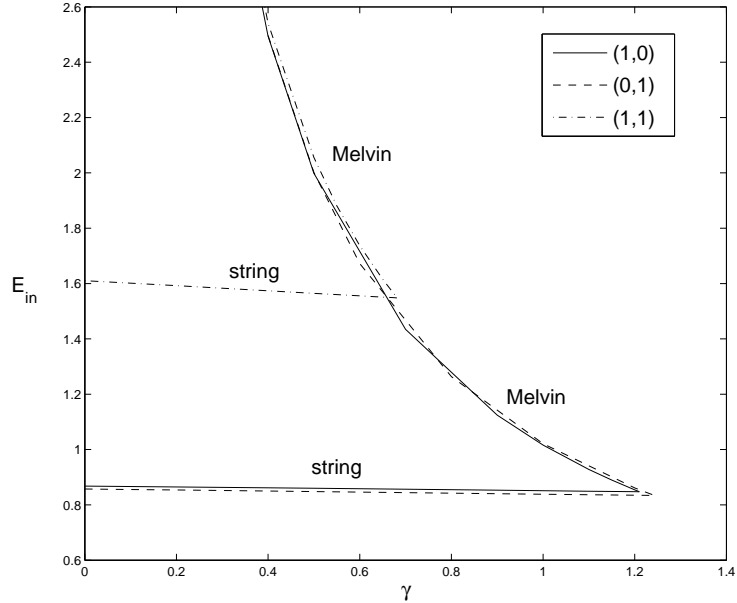} 
\caption{The dependence of the inertial energy $E_{in}$ on the gravitational
coupling $\gamma$ is shown for string and Melvin solutions in the case
of $V_2$ with $\tilde{\beta}_1=\tilde{\beta}_2=1$, $q=g=1$.
Here, we compare the cases $(n,m)=(1,0)$ (solid), $(n,m)=(0,1)$ (dashed) and $(n,m)=(1,1)$ (dotted-dashed).
 }
\label{pot2_critical}
\end{figure}

We have also studied the Melvin solutions in the case of $V_2$. Again, the qualitative
results are very similar to the case with $V_1$. 
In Fig.~\ref{pot2_critical}, we show the dependence of the inertial energy $E_{in}$ on the gravitational coupling $\gamma$ for $\tilde{\beta}_1=\tilde{\beta}_2=1$, $q=g=1$
and three different choices of $(n,m)$. The inertial energy of the string
solutions is always lower than that of the Melvin solutions and the two branches
of solutions meet at $\gamma^{(cr)}$, the maximal value of the gravitational
coupling, beyond which no globally regular solutions exist.
As before, the maximal value of the gravitational coupling decreases with $n+m$ increasing. We include both $(1,0)$ 
and $(0,1)$ solutions, since, as stated above, the do 
differ for the $V_2$ case.

It has been observed previously \cite{bl} that the inertial energy of the Melvin solutions
doesn't depend strongly on the winding number. The reason for this is that the
main contribution to the energy of the Melvin solution comes from the gravitational
field and is thus ``insensitive'' to the actual core structure of the solution. This explains
why the energy of the Melvin solution practically doesn't depend on $\beta_3$ (see Fig.~\ref{fig4}), while it has a strong dependence on the gravitational coupling (see Fig.~\ref{pot2_critical}).
One could see these effects by studying the gravitationally active mass, the so-called Tolman mass $M_{tol}$ \cite{tolman}, which for our definition of the metric is given by \cite{clv}: $M_{tol}\propto
{\rm lim}_{x\rightarrow \infty} (LNN')$. Clearly, for the standard string solutions,
this mass is zero, while for the Melvin solutions it is non-vanishing. The Tolmann mass of the Melvin solutions would then
depend on the windings of the strings (see also the discussions in \cite{bl}).

\section{Conclusions and discussion}

In this paper we have studied the gravitating properties of strings 
arising in field theoretical approaches to (p,q)-superstrings, by minimally
coupling the field theoretical models to gravity. We find that there exist 
globally regular solutions of both the string type (solutions with conical 
deficit) and the Melvin type for a finite domain in parameter space.

The existence of globally regular solutions depends very strongly on the 
(inertial) energy of the underlying string specimen. Therefore, the specific 
choices
of windings of the strings and the string interaction 
(as well as obviously the strength of the gravitational interaction) 
will determine whether regular solutions exists. Due to the interaction
term, the inertial energy of a (p,q) string is lower than the sum of the 
inertial energy of its constituents; thus, the parameter region where
regular strings exist gets enlarged with a stronger interaction term. We also
 found that it is not only the interaction between strings which makes 
the binding stronger, but a stronger gravitational interaction increases
the binding further.

Besides string-like solutions, we have also studied  Melvin solution
for these models. Like their string counterparts, they exist for a finite
range in parameter space, and the range gets enlarged when the interaction
between strings gets stronger. There is some critical value of the parameters
beyond which no regular solution exists. At that critical value, the inertial
energy of the Melvin and the string solution coincide, i.e. the solutions are degenerate. 
For regular solutions
at values of the parameters which are not critical, the Melvin solution
is always heavier than the string solution.

While we have studied the situation where we have straight infinitely long 
strings, it would be very interesting to study the situation where three 
segments of strings meet in a Y junction. The gravitational lensing of 
such a Y junction has been studied for cosmic superstrings using several 
approximations \cite{shlaer}. It would be interesting to see whether we 
can learn something from the
full study of these Y-junctions in field theory, and how they could be 
translated to proper cosmic superstrings.

Another interesting situation would be the following: suppose that we have two strings with winding numbers such that their regular solution exists for 
a given set of parameters. Let us suppose also that the parameters are such 
that
when these two strings interact to form a new bound string, there is no regular
solution of these new string. Then, we would have formed a Y junction in which 
one of the legs is a ``supermassive string'' with a zero of the metric at some
finite value of the radial coordinate. This situation would be worth studying.

There are other extensions of this work which deserve being studied. 
Even though we are dealing with ``extensions'' of the  Abelian Higgs
 model here, string solutions
in non-Abelian models have also been discussed \cite{nona}. 
It would be interesting to study the gravitational effects of the bound 
states in such models.

\begin{acknowledgments}

We acknowledge support from the Marie Curie Intra-European Fellowship
MEIF-CT-2005-009628 (J.U.). This work was partially supported
by the Basque Government (IT-357-07), the Spanish Consolider-Ingenio 2010
Programme CPAN (CSD2007-00042) and  FPA2005-04823 (J.U.).

\end{acknowledgments}


\begin{thebibliography}{99}
\bibitem{polchinski}
  See for instance, J.~Polchinski,  arXiv:hep-th/0412244;   
A.~C.~Davis and T.~W.~B.~Kibble,  Contemp.\ Phys.\  {\bf 46}, 313 (2005).
\bibitem{majumdar} 
  M.~Majumdar and A.~Christine-Davis,  JHEP {\bf 0203}, 056 (2002);  
S.~Sarangi and S.~H.~H.~Tye,  Phys.\ Lett.\  B {\bf 536}, 185 (2002);  
N.~T.~Jones, H.~Stoica and S.~H.~H.~Tye,  Phys.\ Lett.\  B {\bf 563}, 6 (2003).
\bibitem{dvali}
  G.~Dvali and A.~Vilenkin,  JCAP {\bf 0403}, 010 (2004);  
E.~J.~Copeland, R.~C.~Myers and J.~Polchinski,  JHEP {\bf 0406}, 013 (2004);
 A. Ach\'ucarro and J. Urrestilla,  JHEP {\bf 0408}, 050 (2004).

\bibitem{bulk}
C.J.A.P. Martins, Phys. Rev. {\bf D70}, 107302 (2004);
M. Sakellariadou,  JCAP {\bf 0504}, 003 (2005);
E.J. Copeland and P.M. Saffin, JHEP {\bf 0511}, 023 (2005); 
S.-H. Tye, I. Wasserman and M. Wyman, Phys. Rev. {\bf D71}, 103508 (2005);
M.G. Jackson, N.T. Jones and J. Polchinski, JHEP {\bf 10}, 013 (2005);
A. Avgoustidis and E.P.S. Shellard,  Phys.\ Rev.\  {\bf D73}, 041301 (2006);
M. Hindmarsh and P.M. Saffin, JHEP {\bf 0686}, 066 (2006);
E.J. Copeland, T.W.B. Kibble, D.A. Steer, Phys. Rev. Lett. {\bf 97}, 021602 (2006);
E.J. Copeland \etal,   Phys.\ Rev.\  {\bf D77}, 063521 (2008);
A. Avgoustidis and E.P.S. Shellard, arXiv: astro-ph/0705.3395;
H. Firouzjahi, Phys.\ Rev.\  {\bf D77}, 023532 (2008);
R.J. Rivers and D.A. Steer,  arXiv:0803.3968 [hep-th];
N. Bevis and P.M. Saffin,  arXiv:0804.0200 [hep-th].

\bibitem{saffin}
  P.M. Saffin,  JHEP {\bf 0509}, 011 (2005).

\bibitem{rajantie} A. Rajantie, M. Sakellariadou and H. Stoica, JCAP {\bf 0711}, 021 (2007).

\bibitem{salmi} P.Salmi \etal,  Phys.\ Rev.\ {\bf D77}, 041701 (2008).

\bibitem{urrestilla}
  J. Urrestilla and A. Vilenkin,  JHEP {\bf 0802}, 037 (2008).



\bibitem{lyth}
  D.H. Lyth and A. Riotto,  Phys.\ Rept.\  {\bf 314}, 1 (1999).

\bibitem{jeannerot}
  R. Jeannerot, J. Rocher and M. Sakellariadou,  Phys.\ Rev.\ {\bf D68}, 103514 (2003).
  

\bibitem{cmb}
N. Bevis \etal,  Phys. Rev. {\bf D75}, 065015 (2007);  
N. Bevis \etal,  Phys.\ Rev.\ Lett.\  {\bf 100}, 021301 (2008);
N. Bevis \etal,  Phys. Rev. {\bf D76}, 043005 (2007);
J. Urrestilla \etal, arXiv:0711.1842 [astro-ph];
J. Urrestilla \etal, arXiv:0803.2059 [astro-ph].

\bibitem{garfinkle} D. Garfinkle, Phys.\ Rev. \ {\bf D32}, 1323 (1985).

\bibitem{gl} D. Garfinkle and P. Laguna, Phys.\ Rev.\  {\bf D39}, 1552 (1989);
M. E. Ortiz, Phys.\ Rev. \ {\bf D43}, 2521 (1991).

\bibitem{clv} M. Christensen, A.L. Larsen and Y. Verbin, Phys.\ Rev.\ {\bf D60}, 125012 (1999).
\bibitem{bl} Y. Brihaye and M. Lubo, Phys.\ Rev. \ {\bf D62}, 085004 (2000).


\bibitem{deputter}
  A. Ach\'ucarro and R. de Putter,  Phys.\ Rev.\  {\bf D74}, 121701 (2006).


\bibitem{bettencourt}
  L.M.A. Bettencourt and T.W.B. Kibble,  Phys.\ Lett.\  B {\bf 332}, 297 (1994);
  L.M.A. Bettencourt and R.J. Rivers,  Phys.\ Rev.\  {\bf D51}, 1842 (1995).

\bibitem{donaire} 
M. Donaire and A. Rajantie, Phys.\ Rev.\ {\bf D73}, 063517 (2006). 
 
\bibitem{pickles}
M. Pickles and J. Urrestilla,   JHEP {\bf 0301}, 052 (2003); 
Y. Cui, S.P. Martin, D.E. Morrissey and J.D. Wells,  Phys.\ Rev.\  D {\bf 77}, 043528 (2008).


\bibitem{ahu}
A. Ach\'ucarro, B. Hartmann and J. Urrestilla,  JHEP {\bf 0507}, 006 (2005)

\bibitem{composite}
E.R. Bezerra de Mello, Y. Brihaye and B. Hartmann,  Phys.\ Rev.\ {\bf D67}, 045015 (2003);
A. Ach\'ucarro and J. Urrestilla,  Phys.\ Rev.\ {\bf D68}, 088701 (2003);
E.R. Bezerra de Mello,  Phys.\ Rev.\ {\bf D68}, 088702 (2003).


\bibitem{no} H. B. Nielsen and P. Olesen, Nucl.\ Phys.\ B {\bf 61}, 45 (1973).
\bibitem{tolman} R. C. Tolman, Phys.\ Rev.\ {\bf 35}, 875 (1930).
\bibitem{shlaer}
  B. Shlaer and M. Wyman,  Phys.\ Rev.\ {\bf D72}, 123504 (2005);
  R. Brandenberger, H. Firouzjahi and J. Karouby,  Phys.\ Rev.\  {\bf D77}, 083502 (2008).

\bibitem{nona} H. J. de Vega and F. A. Schaposnik, Phys. \ Rev. \ Lett. {\bf 56}, 2564 (1986);
Phys.\ Rev.\ {\bf D34}, 3206 (1986); Y. Brihaye and Y. Verbin, Phys. \ Rev. \  {\bf D77}, 105019 (2008).

\end{thebibliography}
\end{document}